\documentclass[%
aip,
amsmath,amssymb,
reprint,%
]{revtex4-2}

\usepackage{dcolumn}
\usepackage{psfrag}
\usepackage{color}
\usepackage{graphicx,float,amssymb}
\usepackage[bf,loose,normalsize,FIGTOPCAP]{subfigure}
\usepackage{ulem}
\usepackage{mathrsfs} 
\usepackage{color}
\usepackage{graphicx}
\usepackage{natbib}
\usepackage{bm}
\usepackage{amsmath}
\usepackage[colorlinks = true,
linkcolor = blue,
urlcolor  = blue,
citecolor = blue,
anchorcolor = blue]{hyperref}

\usepackage{subfigure}
\usepackage{wrapfig}
\usepackage{amsfonts}

\begin{document}

	\title{Accumulation and alignment of elongated gyrotactic swimmers {in turbulence}}

	\author{Zehua Liu}
	\affiliation{Department of Engineering Mechanics, School of Aerospace Engineering, Tsinghua University, Beijing 100084, China} 
	\author{Linfeng Jiang}
	\email{jianglf17@mails.tsinghua.edu.cn}
	\affiliation{Center for Combustion Energy, Key Laboratory for Thermal Science and Power Engineering of Ministry of Education, Department of Energy and Power Engineering, Tsinghua University, Beijing 100084, China}
	\author{Chao Sun}
	\affiliation{Center for Combustion Energy, Key Laboratory for Thermal Science and Power Engineering of Ministry of Education, Department of Energy and Power Engineering, Tsinghua University, Beijing 100084, China}
	\affiliation{Department of Engineering Mechanics, School of Aerospace Engineering, Tsinghua University, Beijing 100084, China}

	\date{\today}

	\begin{abstract}
		We study the dynamics of gyrotactic swimmers in turbulence, whose orientation is governed by gravitational torque and local fluid velocity gradient. The gyrotaxis strength is measured by the ratio of the Kolmogorov time scale to the reorientation time scale due to gravity, and a large value of this ratio means the gyrotaxis is strong. By means of direct numerical simulations, we investigate the effects of swimming velocity and gyrotactic stability on spatial accumulation and alignment. Three-dimensional Vorono{\"\i} analysis is used to study the spatial distribution and time evolution of the particle concentration. We study spatial distribution by examing the overall preferential sampling and where clusters and voids (subsets of particles that have small and large Vorono{\"\i} volumes respectively) form. Compared with the ensemble particles, the preferential sampling of clusters and voids is found to be more pronounced. The clustering of fast swimmers lasts much longer than slower swimmers when the gyrotaxis is strong and intermediate, but an opposite trend emerges when the gyrotaxis is weak.  In addition, we study the preferential alignment with the Lagrangian stretching direction, with which passive slender rods have been known to align. We show that the Lagrangian alignment is reduced by the swimming velocity when the gyrotaxis is weak, while the Lagrangian alignment is enhanced for the regime in which gyrotaxis is strong.    
	\end{abstract}

	\maketitle

	\section{Introduction}
	Understanding the dynamics and transport of motile particles in a flow is essential for the ecology of aquatic environments. Marine bacteria and phytoplankton live in a world of chemical and physical gradients, and many species actively propel themselves to exploit such heterogeneity \cite{stocker2012marine,Qiu2022}. 
	
	The swimming direction of a micro-organism is influenced by the viscous torque exerted on cells by fluid shear, and by the response to external gradients and biases like gradients of dissolved organic matter and gravity. According to the type of biases, the resulting directed motility is termed chemotaxis \cite{Chakraborty2018}, gyrotaxis \cite{kessler1985hydrodynamic}, and phototaxis \cite{Panda2020,Dervaux2017}.

	In this study, we consider small, elongated, gyrotactic, swimming particles in homogenous isotropic turbulence (HIT). Many motile phytoplankton species are gyrotactic, i.e., their swimming direction results from the competition between shear-induced viscous torque and the stabilizing torque due to bottom-heaviness \cite{pedley1992hydrodynamic}. 
	The stabilizing torque biases their swimming in the vertical direction, inducing an accumulation of cells in well-lit regions near the surface during daylight. Moreover, both single phytoplankton cells and multicellular phytoplankton chains can have elongated shapes, which makes the study of gyrotactic active particles with prolate shapes necessary \cite{smayda2010adaptations}.
	
	The velocity gradients of flows exert viscous torques on microorganisms, which can modify their swimming direction and transportation. Flow can thus affect the spatial distribution of microorganisms, which then determines their encounter rates with prey, predators, and conspecifics \cite{wheeler2019not}. The viscous torques experienced by microorganisms have a shape dependence. While vorticity generates a torque for both spherical and elongated cells, the strain of rate only affects the swimming direction of elongated cells. The additional source in torques for elongated cells causes their different spatial distribution from spherical ones. In laminar flows, spherical cells accumulate in the center of downwelling vertical pipe flows \cite{kessler1985hydrodynamic,Jiang2020}, become trapped in a thin layer of high shear in horizontal shear flows \cite{durham2009disruption,cencini2019gyrotactic}. {Gyrotactic swimmers are found to accumulate due to gravitactic focusing and wall accumulation in channel Poiseuille flows \cite{Wang2021,Wang2022}.}  In turbulent flows, gyrotactic motility can generate small-scale clusters that are dynamically evolving, both when the effect of the fluid acceleration is included or not \cite{durham2013turbulence,de2014turbulent,fouxon2015phytoplankton}. Although both spherical and elongated cells can form clusters in turbulent flows, the extent of the clustering and where cells accumulate in the water column are very different. The clustering of elongated gyrotactic swimmers is shown to be generally weaker than spherical ones, except when the gyrotaxis is very weak \cite{zhan2014accumulation,pujara2018rotations}. While spherical gyrotactic cells collect in downwelling regions \cite{durham2013turbulence}, elongated counterparts preferentially visit either downwelling or upwelling regions, depending on the swimming velocity and the gyrotaxis strength \cite{gustavsson2016preferential,borgnino2018gyrotactic,lovecchio2019chain}. To date, the spatial structures and temporal evolution of clusters of gyrotactic swimmers are still open questions.
	
	Owing to the incompressibility of the fluid, clusters are formed due to the intrinsic swimming of gyrotactic swimmers, with their swimming direction converging within specific regions of the flow. Therefore, the analysis for the alignment is important, which can also influence ocean light climate by altering light scattering \cite{seymour2011microbial}. To describe the alignment of elongated gyrotactic swimmers, one needs a reference frame. In an Eulerian approach, the alignment with the eigenvectors of the strain rate and the fluid velocity has been studied \cite{zhan2014accumulation,pujara2018rotations,borgnino2019alignment}. It has been found that a simpler picture emerges if one describes the local alignment in the reference formed in the Lagrangian framework \cite{ni2014alignment,hejazi2017emergent,cui_dubey_zhao_mehlig_2020}. The physical image of the Lagrangian reference frame is suggested as below. An infinitesimal, spherical fluid element deforms, over time, into a tri-axial ellipsoid as it is stretched by the turbulent flow, whose dynamic evolution can be naturally described by following its trajectory. After a sphere is distorted into an ellipsoid, the orientations of the three principal axes form an orthogonal coordinate system, which is in accordance with the eigenvectors of the left Cauthy-Green strain tensor \cite{ni2014alignment}. The Lagrangian direction is defined to be the orientation of the longest principal axis of this ellipsoid, implying the strongest stretching along this direction. However, the orientation behavior of active particles with respect to the Lagrangian stretching direction has not been addressed.

	In this study, we focus on the relationship between the clustering of gyrotactic particles and the flow, using the three-dimensional Vorono{\"\i} analysis. The Vorono{\"\i} method allows us to track the concentration of particles in the Lagrangian framework. We examine the p.d.f.s of the Vorono{\"\i} volume, the locations where clusters form, and the lifetime of the clusters. The remaining paper is organized as follows. In Sec.~\ref{sec:method}, we briefly describe the background, explaining the model equations for gyrotactic self-propelled particles and the numerical method we use. The definitions of three-dimensional Vorono{\"\i} tessellation and Lagrangian stretching are also discussed in this section. In Sec.~\ref{sec:results}, we present results on the accumulation and alignment. The conclusions, Sec.~\ref{sec:conclusions}, summarize the main finding of this study.

	\section{Method \label{sec:method}}
	\subsection{Model equations}
	The turbulent flow is governed by the incompressible Navier-Stokes equations (NSE) driven by an external random large-scale statistically homogeneous and isotropic force with constant in-time global energy input:
	
	\begin{eqnarray}
		\frac{\partial {\boldsymbol u}}{\partial t}  + \boldsymbol{u}\cdot {\nabla}   \boldsymbol{u} &=& -   \rho^{-1}{\nabla} p + \nu\ \nabla^2   \boldsymbol{u} + \boldsymbol{f} , \label{NS}\\
		{\nabla}  \cdot  \boldsymbol{u} &=& 0. \label{Imcompress}
	\end{eqnarray}
	Here, $\boldsymbol{u}(\boldsymbol{x},t)$ is the velocity field, $p(\boldsymbol{x},t)$ is the pressure, $\nu$ is the kinematic viscosity, and $\rho$ is the liquid density. The vector $\boldsymbol{f}$ refers to the external large-scale body force needed to sustain statistically stationary turbulence. {The force $\boldsymbol{f}$ to stir turbulence is modulated by means of a sum of sine waves with small wavenumbers and applied at each position and at each time step. The phases of sine waves are evolved in time by means of a stochastic process to ensure a homogeneous and isotropic stirring \cite{Perlekar2012}.} The turbulent intensity of the flow is characterised by a dimensionless parameter, the Taylor-Reyolds number, $Re_\lambda=u_{rms}\lambda/\nu$, where $u_{rms}=\sqrt{\langle u_iu_i\rangle^{v,t}/3}$ is the root-mean-square (r.m.s.) velocity of the flow ($\langle \ldots \rangle^{v,t}$ denotes here volume and time average), $\lambda=\sqrt{15\nu u_{rms}^2/\epsilon}$ is the Taylor length scale and $\epsilon = (\nu/2) \Sigma_{i,j}\langle(\nabla_i u_j + \nabla_j u_i)^2 \rangle^{v,t}$ is the mean global energy dissipation rate.
	
	A gyrotactic swimmer is modeled as an axisymmetric ellipsoid that swims along with its long axis direction $\boldsymbol{p}$ with a constant swimming speed $v_s$. The total particle velocity is the sum of the swimming velocity  $v_s\boldsymbol{p}$ and the flow velocity $\boldsymbol{u}$ at the particle location. Assuming the swimmers are small compared to the dissipative scale of the turbulent flow and neutrally-buoyant, particle trajectories, $\boldsymbol{x}(t)$, and swimming directions, $\boldsymbol{p}$ can be integrated using the following equations \cite{Jeffery1922,kessler1985hydrodynamic,pedley1992hydrodynamic}:
	\begin{eqnarray}
		\frac{\mathrm{d}\boldsymbol x}{\mathrm{d} t} &=& \boldsymbol{u}(\boldsymbol{x}(t),t)+v_s\boldsymbol p,\label{motion}\\
		\frac{\mathrm{d} \boldsymbol p}{\mathrm{d} t} &=&\frac{1}{2B}[\boldsymbol{z}-(\boldsymbol{z} \cdot \boldsymbol{p})\boldsymbol{p}]+\boldsymbol{\Omega} \boldsymbol{p} +\Lambda \left( \boldsymbol S\boldsymbol{p} - (\boldsymbol{p}\cdot \boldsymbol S\boldsymbol{p})  \boldsymbol{p} \right),\quad \label{orientation}
	\end{eqnarray}
	In the above equations, $\boldsymbol{z}$ is the vertically upward direction, 
	$\boldsymbol{\Omega}=\frac{1}{2}(\nabla \boldsymbol{u}-(\nabla\boldsymbol{u})^\mathrm{ T })$ and $\boldsymbol{S}=\frac{1}{2}(\nabla \boldsymbol{u}+(\nabla\boldsymbol{u})^\mathrm{ T })$ are rotation and strain rate tensor, respectively, $\Lambda=(\alpha^2-1)/(\alpha^2+1)$ is the particle eccentricity, $\alpha$ is the particle aspect ratio, which is defined as the ratio the length $l$ to the diameter $d$ of the particle. {$B$} is the gyrotactic reorientation timescale, which measures how long it takes for a perturbed cell to return to a vertical direction in a quiescent flow. {According to the bottom heaviness, the reorientation time scale can be computed using $B=\nu \alpha_\perp/2gh$, where $\nu$ is the kinematic viscosity of the fluid, $\alpha_\perp$ is a shape-dependent dimensionless coefficient that measures the viscous resistance of the tumbling rotation of a particle, $h$ is the distance between the center of gravity and buoyancy, and $g$ is gravity\cite{lovecchio2019chain, pedley1992hydrodynamic}. Therefore, a strong gravity effect or equivalently a strong bottom-heaviness-induced gravitational torque is corresponding to a small $B$.} The three terms on the right-hand side of (\ref{orientation}) reflect that the tumbling rate of an elongated gyrotactic swimmer is governed by the {gravitational} torque, and the torque of the ambient flow through the fluid's vorticity and strain of rate, respectively. In what follows, only particles with $\alpha=20$ will be considered, corresponding to $\Lambda=0.995$. This choice of particle aspect ratio is based on two considerations. First, the individual cell of many phytoplankton species is elongated, often with large aspect ratios \cite{https://doi.org/10.4319/lom.2007.5.396}.    Secondly, the chain formation of phytoplankton when daughter cells remain attached to one another after cell division leads to large aspect ratios. Considering the chain formed by individual spheric particles, $\Lambda$ rapidly saturates with the chain length. Chains of 2, 4, 8, 16, and 32 cells are corresponding to $\Lambda=0.600, 0.882, 0.969, 0.992, 0.998$ \cite{lovecchio2019chain}. This large aspect ratio ($\Lambda$ approaches 1) has also been adopted by many previous studies to investigate the clustering behavior of prolate particles \cite{borgnino2018gyrotactic,lovecchio2019chain}. The suspension is assumed to be diluted, so the interactions between particles are neglected. For studies about particle interaction, we refer to Ref.~\onlinecite{Bechinger2016,Benfenati2013,Sun2006}.

	For elongated gyrotactic swimmers herein, two dimensionless parameters control the particle's fate:
	\begin{eqnarray}
		\Phi=v_s/u_{\eta},
		\label{phi}\\
		\Psi=\tau_{\eta}/B.
		\label{psi}
	\end{eqnarray} 
	
	First, the swimming number $\Phi$ in (\ref{phi}) measures the swimming velocity $v_s$ relative to the Kolmogorov velocity $u_{\eta}=(\nu\epsilon)^{1/4}$. Second, the stability number $\Psi$ in (\ref{psi}) is defined to be the ratio of Kolmogorov time $\tau_{\eta}=(\nu/\epsilon)^{1/2}$ to the gyrotactic reorientation time $B$, as a predictor of a particle's ability to maintain vertical migration. A small gyrotactic reorientation times $B$ implies that the particle can quickly orient upward again after a perturbation. Therefore, $\Psi$ parametrizes the importance of directional swimming with respect to the ambient flow overturning. $\Psi\gg1$ means strong stability, producing a swimming direction that is very stable and uniformly upwards. Conversely, $\Psi\ll1$ reflects that the gyrotactic effect is negligible, and the swimming direction is completely determined by the ambient flow.
	
	\begin{figure}[!ht]
		\begin{center}
			\includegraphics[width=0.8\columnwidth,trim=0 30 0 0,clip]{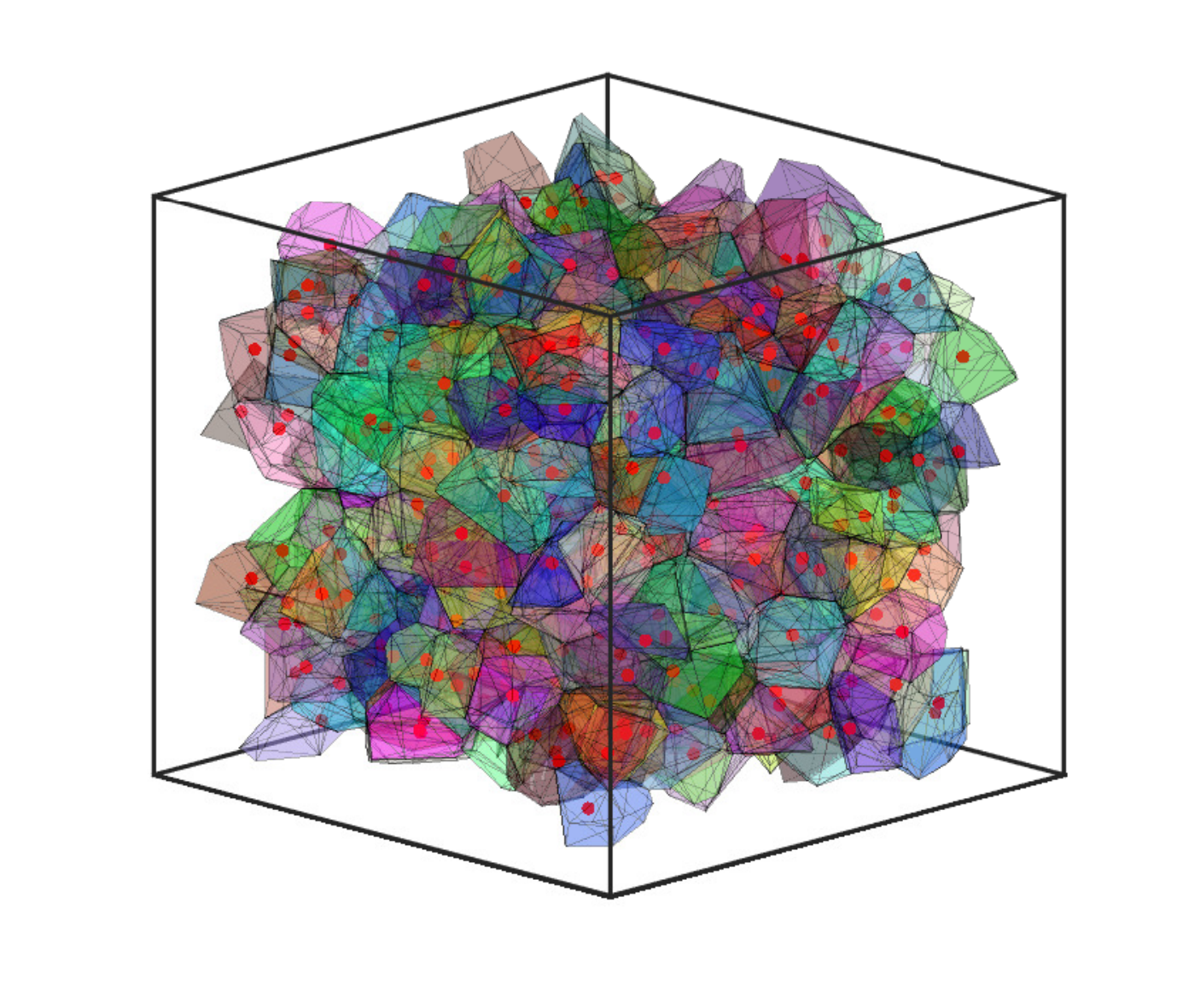}
			\caption{Visualization of a three-dimensional Vorono{\"\i} tessellation. Particles are represented by red points. Each Vorono{\"\i} cell is a polyhedron assigned to a particle at its center.}
			\label{fig:fig1}				
		\end{center}
	\end{figure}
	
	\subsection{Three-dimensional Vorono{\"\i} analysis}
	
	Previous studies have quantified the clustering of gyrotactic active particles by employing different approaches, such as pair correlation function \cite{zhan2014accumulation}, clustering index \cite{borgnino2018gyrotactic}, and correlation dimension \cite{durham2013turbulence,de2014turbulent}. In the above studies, these methods do not provide information on the exact locations where the strongest clustering occurs or the Lagrangian evolution of the clusters. An alternative mathematical tool, the Vorono{\"\i} tessellation, can be used to study the clustering, which provides more information about the clustering \cite{tagawa2012three}. As shown in Figure \ref{fig:fig1}, the Vorono{\"\i} diagram is a unique decomposition of the three-dimensional space into independent cells associated with each particle. One Vorono{\"\i} cell is defined as the ensemble of points that is the closest to the particle position compared to any other. From the definition of the Vorono{\"\i} diagrams, the volume of a Vorono{\"\i} cell is the inverse of the local particle concentration. Vorono{\"\i} volumes can be used to measure the concentration of each individual particle present in the field, and this feature makes Lagrangian tracking of concentration along particle trajectories possible \cite{Monchaux2012}. \citet{ferenc2007size} have found that a $\Gamma$-distribution can adequately describe the probability density functions (p.d.f.s) of the Vorono{\"\i} volumes normalized by the mean volume for randomly distributed particles. In the three-dimensional case, the $\Gamma$-distribution has the following prefactor and exponent:
	\begin{equation}
		f(x)=\frac{3125}{24}x^4\text{exp}(-5x).
		\label{Gamma}
	\end{equation}
	Here {$x$} is the Vorono{\"\i} volume normalized by the mean volume. Particles that have a p.d.f. that deviates the $\Gamma$-distribution, indicate that the combined effect of flow and characteristics of particles can affect the spatial distribution. For those particles at the border of the domain, there may not be enough other particles surrounding them. These Vorono{\"\i} cells are ill-defined, and thus not considered for the analyses.

	\subsection{Lagrangian stretching direction}
	In this part, we briefly discuss how the Lagrangian stretching direction is obtained. For more details about this Lagrangian reference frame, we refer to Ref.~\onlinecite{chadwick1999continuum,ni2014alignment,cui_dubey_zhao_mehlig_2020}. In order to characterize the deformation experienced by the fluid element, we first define a deformation tensor $\boldsymbol {F}(t)$ as the solution of the differential equation: 
	\begin{equation}
		\frac{d}{dt}\boldsymbol{F}(t)=\boldsymbol{A}(\boldsymbol{x},t)\boldsymbol{F}(t).
		\label{deformation}
	\end{equation}
	Here $\boldsymbol{A}(\boldsymbol{x},t)$ is the instantaneous velocity gradient. We obtain $\boldsymbol {F}(t)$ by integrating (\ref{deformation}) using the velocity gradient along the particle trajectories, with initial condition $\boldsymbol{F}(0)=\mathbf{I}$ (identity matrix). For a non-active particle, the particle velocity is always the same as the background fluid velocity at the particle position. In this case, the physical image of $\boldsymbol {F}(t)$ is suggested as below. Consider a fluid element that moves with a non-active particle, and this can be done because they have the same velocity and thus the same trajectory. This infinitesimal initial spherical fluid element is stretched, over time, into an ellipsoid, due to the deformation effect exerted by the fluid velocity gradient. And the deformation tensor characterizes the fluid element deformation in a Lagrangian way \cite{parsa2011rotation,wilkinson2011emergent}. 
	
	The velocity of a motile particle, $\boldsymbol{u}(\boldsymbol{x},t)+v_s\boldsymbol{p}$, is different from the fluid element velocity at the particle position $\boldsymbol{u}(\boldsymbol{x},t)$. We remark that this time there is no real fluid element following a swimming particle. However, by integrating (\ref{deformation}) the velocity gradient $\boldsymbol{A}(\boldsymbol{x},t)$ along a swimmer trajectory, an imaginary initial spherical deformable element is used to follow the swimming particle and record the stretching of the ambient flow. Although there is a slight difference between the meaning of the deformation tensor between non-active and motile particles, in both these two cases, the deformation tensor is a useful intermediate quantity to form a Lagrangian reference frame at the particle position and reflect the accumulative stretching on particles exerted by the ambient flow.
	
	The left Cauthy-Green tensor $\boldsymbol{C}^{(L)}$ is then obtained from the inner product of $\boldsymbol{F}$ with itself:
	\begin{equation}
		\boldsymbol{C}^{(L)}=\boldsymbol{FF}^T.
	\end{equation}    
	The eigenvalues of $\boldsymbol{C}^{(L)}$ are denoted with $\Lambda_i (i=1,2,3)$, and eigenvectors $\boldsymbol{e}_{Li}$ correspondingly. 
	In the case of an incompressible flow, $Tr\boldsymbol A = 0$ and the eigenvalues of the left Cauchy–Green tensor satisfy $\Lambda_1(t)\Lambda_2(t)\Lambda_3(t)=1$.
	The largest eigenvalue, $\Lambda_1>1$, indicates extension, and the corresponding eigenvector $\boldsymbol{e}_{L1}$ is called the Lagrangian stretching direction. The smallest eigenvalue, $\Lambda_3<1$, indicates contraction in the $\boldsymbol{e}_{L3}$ direction. As for the intermediate eigenvalue $\Lambda_2$, it can either be greater or less than $1$, reflecting either stretching or compressing in the $\boldsymbol{e}_{L2}$ direction. 
	
	\subsection{Numerical simulations}
\begin{figure}
		\begin{center}
			\includegraphics[width=1\columnwidth]{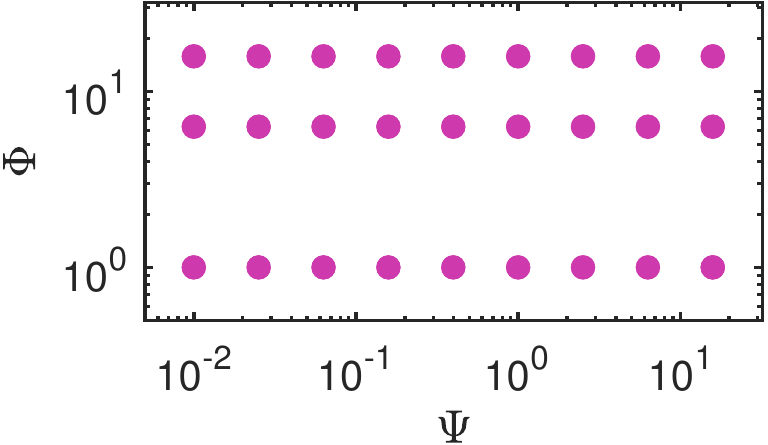}
			\caption{Explored parameter space of $\Psi$ and $\Phi$ for our numerical study of gyrotactic swimmers.}
			\label{fig:fig2}				
		\end{center}
	\end{figure}

	The NSE here is numerically simulated by using a Lattice Boltzmann Method code, the {ch4-project} \cite{calzavarini2019eulerian}, which has been extensively employed in the studies of Lagrangian tracers and point-like particle dynamics in turbulence \cite{jiang2020rotation,Calzavarini2020,Jiang2021}. {More specifically, recent studies of rotational dynamics of inertialess ansiotropic particles \cite{jiang2020rotation} and gyrotactic particles \cite{Jiang2021} in turbulence have validated the particle dynamics model and the code through comparison with experimental results.} This code uses a tri-linear scheme for the Lagrangian-Eulerian frame interpolations.
	We refer to Ref.~\onlinecite{calzavarini2019eulerian} for the details of the numerical method.
	For the present simulations, a resolution of $256^3$ grid is used for the HIT flow at $Re_\lambda\approx60$.
	The boundary conditions are periodic in all directions of the three-dimensional cubic simulation domain. As indicated in Figure \ref{fig:fig2}, a wide range of particle parameters has been explored, with $\Phi=1,~6.3,~\text{and}~16$ and $\Psi\in[0.01, 16]$. For each parameter set, $5\times10^4$ Lagrangian trajectories are 
	seeded in the flow and the duration is about 20 large-eddy turnover times. The deformation tensor $\boldsymbol {F}(t)$ is integrated from the DNS data using a fourth-order Runge-Kutta scheme according to (\ref{deformation}). The random initial conditions are adopted for the particle positions and orientations. All analyses are performed after particles have reached a statistically stationary state.

	\section{Results and discussion}\label{sec:results}
	We aim to address the following problems: (i) How the clustering is affected by the swimming number $\Phi$ and the stability number $\Psi$? In particular, how to characterize the extent of clustering based on the three-dimensional Vorono{\"\i} analysis? (ii) How the clustering is related to the flow structures? This question has two perspectives. From the space perspective, what are the regions that particles accumulate in? From the time perspective, how long do particles exist in an aggregated state?
	
	\subsection{Particle accumulation}\label{sec:particle accumulation}
	
	\begin{figure}
		\begin{center}
			\includegraphics[width=1\columnwidth]{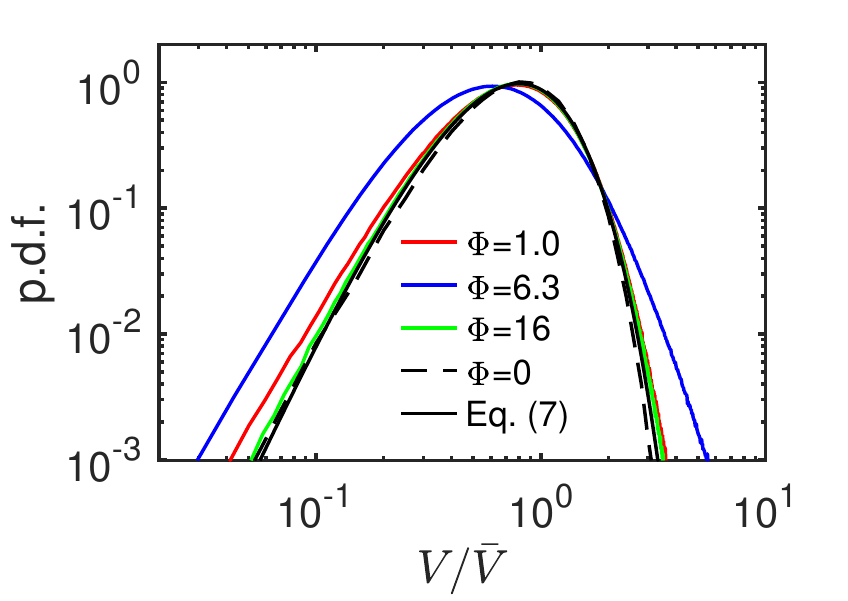}
			\caption{The p.d.f.s for the normalized Vorono{\"\i} cell volumes of fast ($\Phi=16$), moderate ($\Phi=6.3$), and slow swimmers ($\Phi=1.0$) at $\Psi=0.01$, as compared with the distributions obtained from non-active particles (dashed line) and randomly distributed particles (\ref{Gamma}). All three types of motile particles show clustering, whereas non-active particles remain randomly distributed.}
			\label{fig:fig3}				
		\end{center}
	\end{figure}

	\begin{figure*}
		\begin{center}
			\includegraphics[width=2\columnwidth]{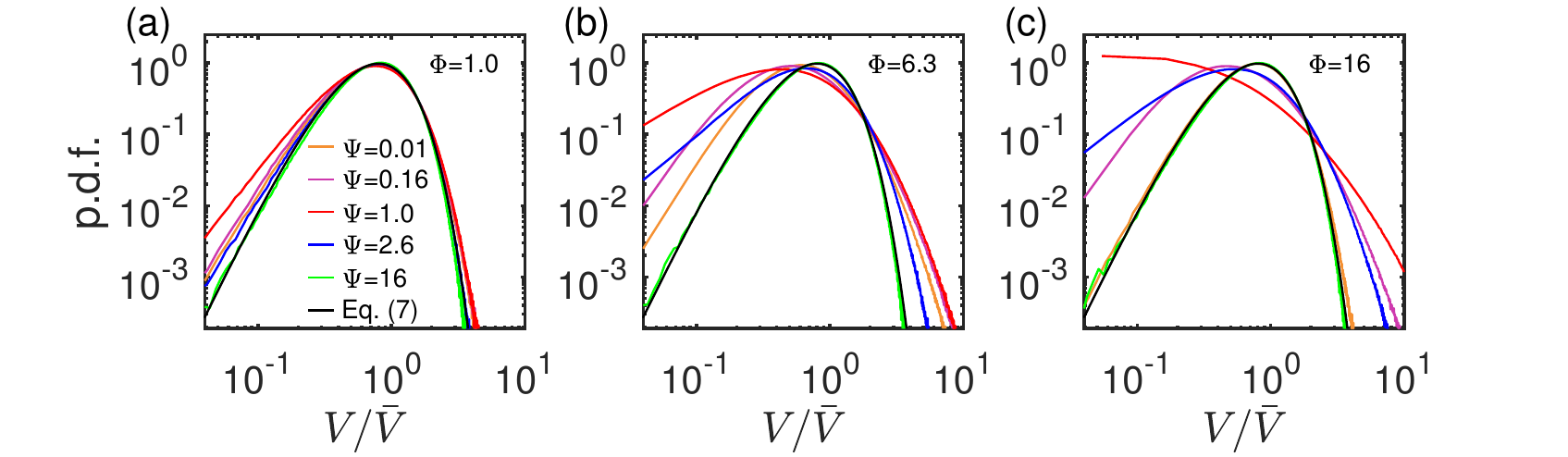}
			\caption{The normalized Vorono{\"\i} volume p.d.f.s for (a) slow swimmers ($\Phi=1.0$), (b) moderate swimmers ($\Phi=6.3$), (c) fast swimmers ($\Phi=16$). The results of different $\Psi$ are plotted for each type of particle.  No preferential concentration is observed in the gravity-dominated case ($\Psi=16$), whereas rod-like slow and moderate swimmers maintain observable clustering even when the gravity effect is nearly negligible ($\Psi=0.01$). At intermediate values of $\Psi$, the clustering is most pronounced.}
			\label{fig:fig4}				
		\end{center}
	\end{figure*}
	
	As shown in Figure \ref{fig:fig3}, the p.d.f.s of the Vorono{\"\i} volume ($\mathscr{V}$) normalized by the global mean value ($\mathscr{\bar{V}}$), $\mathscr{V}/\mathscr{\bar{V}}$, is plotted for particles of different $\Phi$ at $\Psi=0.01$ to investigate the effect of swimming velocity on the clustering. The gyrotactic effect can be considered to be negligible at $\Psi=0.01$. 
	For non-active particles ($\Phi=0$), no preferential concentration is expected, with the p.d.f. agreeing well with that of randomly distributed particles. 
	In contrast, for active particles, we observe different behavior compared to the randomly distributed particles.  
	For all three types of active particles at $\Psi=0.01$, the possibilities of attaining either small or large normalized Vorono{\"\i} volumes are larger than that of randomly distributed particles, which are corresponding to clusters and voids in the domain. 
	For slow swimmers, the effect of the swimming velocity on the particle dynamics is not significant, which leads to a slight increase in the probability of small Vorono{\"\i} volumes. However, when the particle dynamics is dominated by the swimming velocity ($\Phi=16$), particle accumulation becomes less evident as the particles may tend not to be trapped due to high swimming velocity. The strongest clustering is observed for moderate swimmers ($\Phi=6.3$). As studied in Ref.~\onlinecite{pujara2018rotations}, anisotropic swimmers without gyrotaxis accumulate in regions of lower fluid vorticity. In the swimming number range they covered, $\Phi\in[0, 5]$, the extent of accumulation monotonically increases with the particle swimming velocity. This is consistent with the variation of the p.d.f.s at $\Psi=0.01$ from $\Phi=1$ to $\Phi=6.3$.

	\begin{figure}
		\begin{center}
			\includegraphics[width=1\columnwidth]{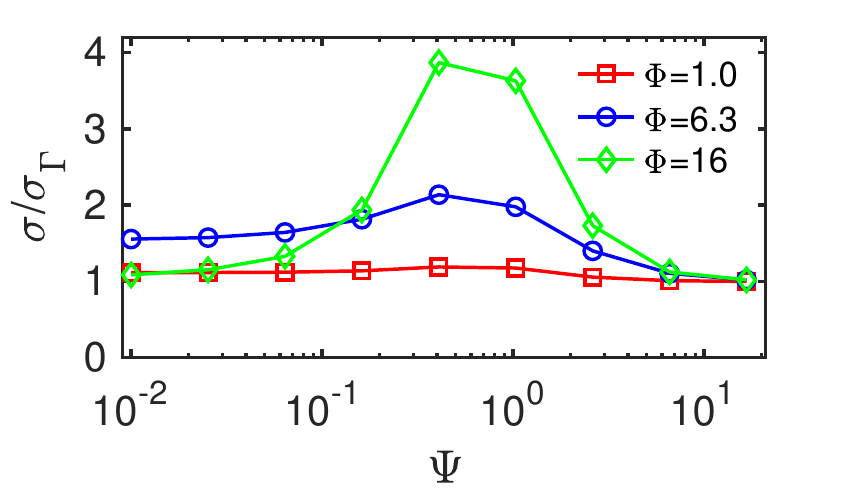}
			\caption{
				The normalized standard deviation of Vorono{\"\i} cell volumes, $\sigma/\sigma_\Gamma$, as a function of $\Psi$ at $\Phi$ of 1.0, 6.3, and 16.}
			\label{fig:fig5}       
		\end{center}
	\end{figure}
	
	Figure~\ref{fig:fig4} shows the p.d.f.s of the normalized Vorono{\"\i} volume for different $\Psi$ at $\Phi$ of 1.0, 6.3, and 16.
	The gravity-induced torque further affects the particle dynamic so as to change the particle accumulation. 
	It turns that, for all swimming numbers, gyrotaxis enhances the probability of finding either small or large Vorono{\"\i} volumes and the influence becomes highest when the flow time scale and the gyrotactic reorientation time are comparable ($\Psi=1.0$). More importantly, this phenomenon even exists when the swimming velocity is dominated ($\Phi=16$), whereas no particle accumulation is observed for non-gyrotactic fast swimmers as shown in Figure~\ref{fig:fig3}. 
	Furthermore, the overall extent of particle accumulation increases with the swimming number. When the gyrotactic effect is dominated ($\Psi=16$), the particles are compelled to point upward, which leads to random distributions similar to $\Gamma$-distribution (\ref{Gamma}).
	
	\begin{figure*}
		\begin{center}
			\includegraphics[width=1.4\columnwidth,height=6cm,trim=70 0 70 0,clip]{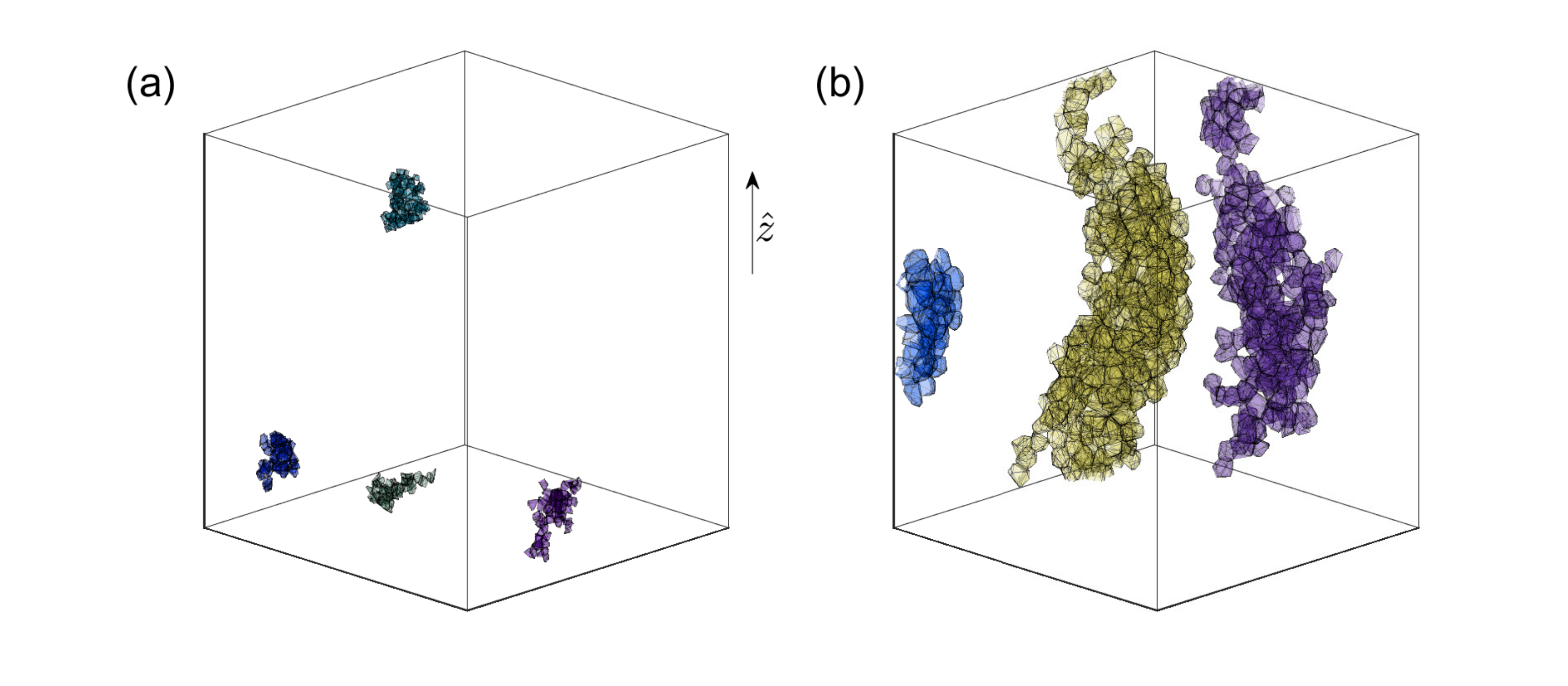}
			\caption{Several largest clusters and voids in a snapshot of particles at $\Phi=16$ and $\Psi=0.40$, with the Vorono{\"\i}-cell faces colored uniformly per cluster. The particles outside these clusters and voids have been omitted from the figure. Panel (a) and (b) shows the clusters and voids, respectively.}
			\label{fig:fig6}        
		\end{center}
	\end{figure*}
	
	\begin{figure*}
		\begin{center}
			\includegraphics[width=2\columnwidth]{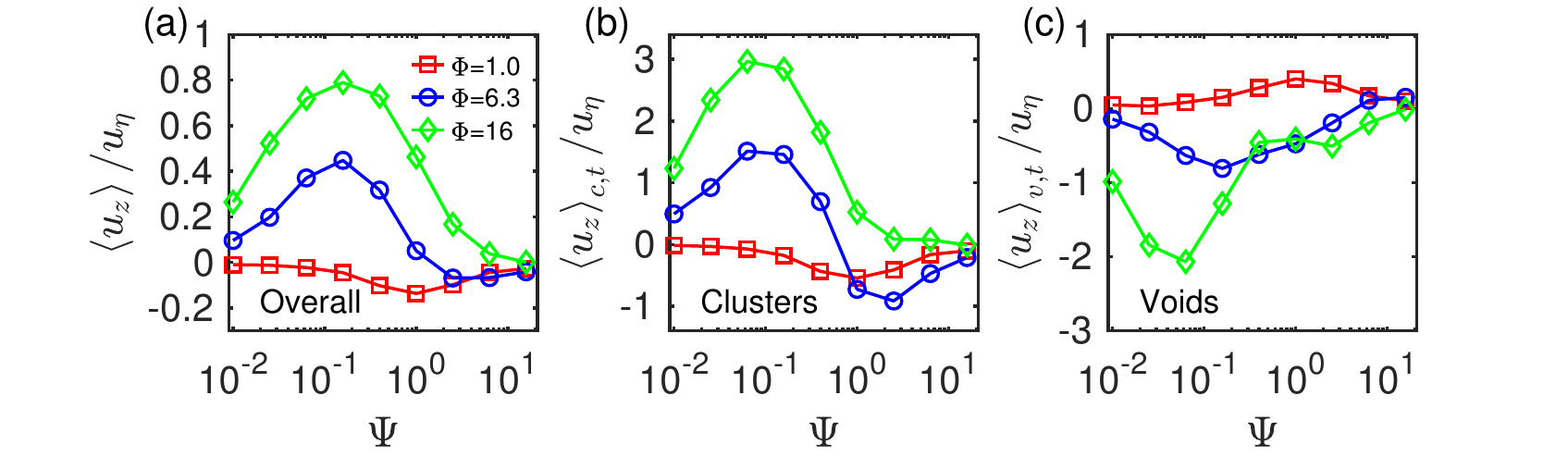}
			\caption{The mean vertical fluid velocity at the positions of (a) overall particles, (b) clusters, (c) voids.  
			}
			\label{fig:fig7}				
		\end{center}
	\end{figure*}
	
	The extent of particle accumulation can be further quantified by the standard deviation of normalized Vorono{\"\i} volumes, $\sigma/\sigma_{\Gamma}$, where $\sigma_{\Gamma}$ is the standard deviation of Vorono{\"\i} volumes for randomly distributed particles. This quantity has been adopted by other studies on inertial particle clustering in turbulence \citep{monchaux2010preferential,tagawa2012three}. 
	A value of $\sigma/\sigma_{\Gamma}=1$ represents that no clustering can be observed, and high values of this term reveal the existence of high and low concentration events. We remark that $\sigma/\sigma_{\Gamma}$ is dependent on the number of particles in the domain. To test this particle number dependence, we calculated $\sigma/\sigma_{\Gamma}$ for a reduced number of particles by randomly removing some particles. The value of the indicator $\sigma/\sigma_{\Gamma}$ can be biased by this subsampling procedure, and it is reduced by nearly $20\%$ when $80\%$ of the $50000$ particles in total belonging to the group of $\Psi=1$ and $\Phi=16$ are randomly removed from the domain.  
	However, $\sigma/\sigma_{\Gamma}$ remains a good indicator for a fixed number of particles, with larger values indicating more clustering. As seen in Figure \ref{fig:fig5}, for the range of explored stability numbers (spanning from 0.01 to 16), the standard deviation of the normalized Vorono{\"\i} volumes of all three types of swimmers shows a non-monotonic trend, with the indicator value peaking at $\Psi=O(1)$. Under the intermediate gyrotaxis, fast swimmers exhibit the strongest clustering, reaching a value of 3.9. The clustering result has a consistent trend with that of Ref.~\onlinecite{zhan2014accumulation}, which evaluated the clustering extent by the slope of the radial distribution function. {Similar clustering extent trend has also been presented through the correlation dimension $D_2$ \cite{borgnino2018gyrotactic}, which rules the small-scale scaling behaviour of the probability to find a pair of cells with distance less than $r$, i.e., $p_2(r)\sim r^{D_2}$ when $r\to 0$.}
	
	\subsection{Clusters and preferential sampling }\label{sec:flow}
	
	In this section, we aim to study how the spatial distribution of particles is related to the turbulent structures. The particle parameters affect not only the extent of the clustering but, importantly, also the locations of the clustering. It has been found that rod-like gyrotactic swimmers preferentially visit either downwelling or upwelling regions depending on the $\Psi$ and $\Phi$ \citep{borgnino2018gyrotactic,lovecchio2019chain}. Therefore, we investigate this issue by examing the vertical fluid velocity sampled by particles and their connections with the Vorono{\"\i} cell volumes. In order to identify where clusters and voids form, we denote $\mathscr{V}_f$ as the Vorono{\"\i} volume which a fraction $f$ of Vorono{\"\i} cells are smaller than, i.e. $P(\mathscr{V}<\mathscr{V}_f)=f$. Here, normalized Vorono{\"\i} volumes smaller than $\mathscr{V}_{0.05}$ are defined as belonging to clusters. Equivalently, Vorono{\"\i} cells larger than $\mathscr{V}_{0.95}$ are defined as belonging to voids. In other words, Vorono{\"\i} cells whose volume is below the threshold defined by $\mathscr{V}_{0.05}$ belong to clusters, while Vorono{\"\i} cells above $\mathscr{V}_{0.95}$ belong to voids. Figure \ref{fig:fig6} shows an example of several identified largest clusters and voids in one snapshot from the simulation. It appears that these Vorono{\"\i} cells that possess extreme sizes tend to be connected in groups of various sizes and shapes. The sizes of voids are naturally larger than that of the clusters, due to the larger sizes of connected Vorono{\"\i} cells. And the elongation of the shapes of voids in the vertical direction is observed.
	It is now tempting to ask these questions: where do the overall particles ($\mathscr{V}_0<\mathscr{V}<\mathscr{V}_1$) preferentially sample? where do clusters ($\mathscr{V}<\mathscr{V}_{0.05}$) and voids ($\mathscr{V}>\mathscr{V}_{0.95}$) preferentially form?

	First, we focus on the overall preferential sampling. We use the non-dimensional vertical fluid velocity (normalized by the Kolmogorov velocity) averaged over the positions of all particles, $\left \langle u_z \right \rangle /u_{\eta}$, to show this preferential sampling of fluid velocities, where $\langle \ldots \rangle$ denotes the ensemble average over the locations of overall particles ($\mathscr{V}_0<\mathscr{V}<\mathscr{V}_1$) and over time. Similarly, in what follows, $\langle \ldots \rangle_{c,t}$ and $\langle \ldots \rangle_{v,t}$ denote the time average over the locations of clusters ($\mathscr{V}<\mathscr{V}_{0.05}$) and voids ($\mathscr{V}>\mathscr{V}_{0.95}$) respectively.
	In Figure \ref{fig:fig7}(a), the different curves are plotted for $\left \langle u_z \right \rangle/u_{\eta}$ of three types of swimmers at different values of $\Phi$ as a function of the stability number $\Psi$. There is a clear difference among slow, moderate, and fast swimmers. Slow swimmers, at all $\Psi$ in the range we have covered, tend to stay in the downwelling flow, whereas fast swimmers consistently preferentially visit upwelling regions. Moderate swimmers, preferentially sample regions with a positive vertical velocity at small $\Psi$, but regions with a negative vertical velocity at large $\Psi$. {Field observation shows that the dinoflagellate \textit{Cochlodinium polykrikoides} tends to be in long chain groups of $\ge 6$ cells before sunrise, whereas the reverse is observed after that with the densities of short chain groups of $\le5$ cells increases \cite{JongGyuPark2001}. The study notes that this is a ecological strategy through which the population can use the sufficient propulsion produced by more cells to swim fast to move upward before sunrise. After arrival in the surface layers, the shape of short chain provides a high area-to-volume ratio for them to capture energy efficiently. The tendency of fast swimmers to preferentially sample in upwelling regions in Figure 7(a) implies that a long chain can help the vertical upward migration not only through fast swimming speed but also through a turbulent upwelling water column.}  
	
	Here, we answer the question of where clusters and voids form. Figure \ref{fig:fig7}(b, c) show the average non-dimensional vertical fluid velocity sampled by clusters $\left \langle u_z \right \rangle _{c,t}/u_{\eta}$ and voids $\left \langle u_z \right \rangle _{v,t}/u_{\eta}$ respectively. The curves in Figure \ref{fig:fig7}(b) for clusters roughly show a similar trend with the overall case, reflecting that the cluster locations qualitatively agree with locations that overall particles preferentially visit. The opposite trend is observed for voids in Figure \ref{fig:fig7}(c), reflecting that they form at regions where the overall particles tend to avoid. From a quantitative perspective, the mean velocity sampled by the clusters is larger than that of overall particles for most particle parameters (namely $\Phi$ and $\Psi$). This demonstrates that the clusters preferentially form at higher vertical velocity flow regions compared with the overall preferential sampling. Particles that preferentially sample in upwelling regions or downwelling regions, respectively, form clusters in upflows or downflows with more extreme vertical velocity. The voids also tend to form at higher velocity vertical flows compared with the overall particles, though the vertical flow is in the opposite direction to that of the overall particles.
	
	\subsection{Vorono{\"\i} Lagrangian autocorrelation} \label{sec:lagcorr} 
	
	\begin{figure}
		\begin{center}
			\includegraphics[width=1\columnwidth]{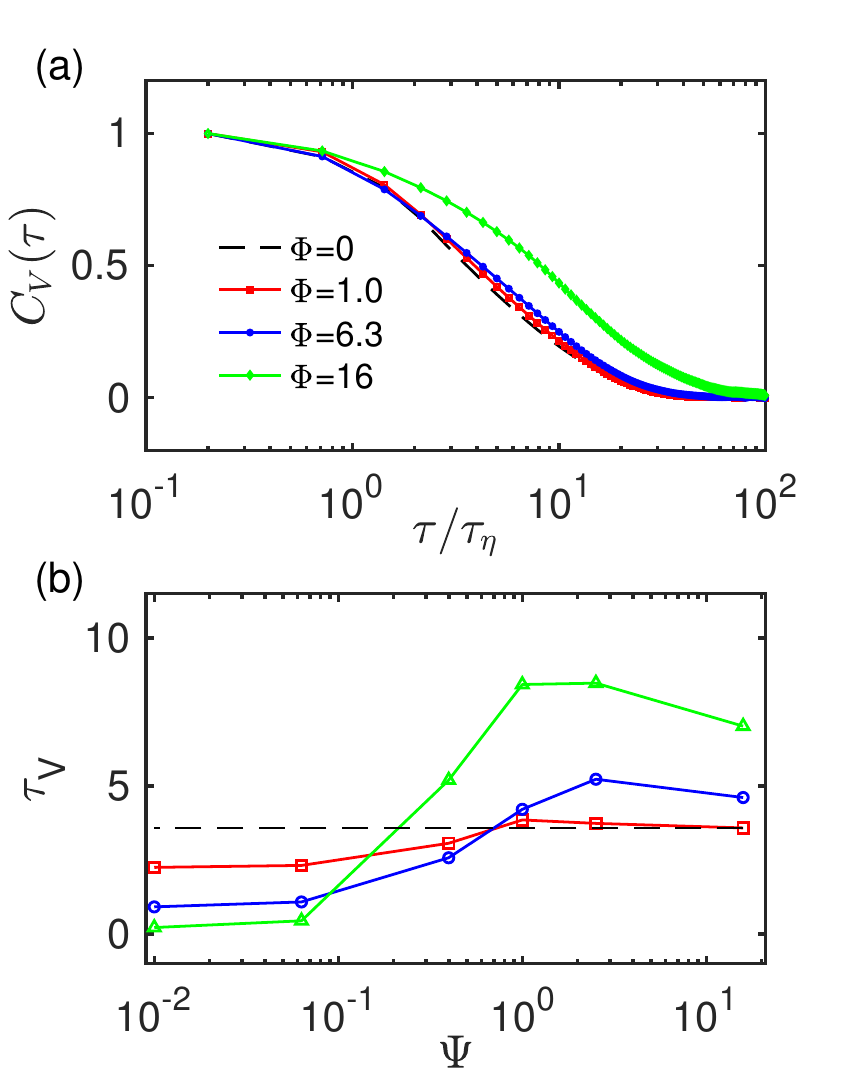}
			\caption{Lagrangian Vorono{\"\i} analysis for fast ($\Phi=16$), moderate ($\Phi=6.3$), slow swimmers ($\Phi=1.0$) and non-active particles (dashed line). (a) Temporal autocorrelation functions of Vorono{\"\i} volumes at $\Psi=1$. (b) Decorrelation time.}
			\label{fig:fig8}				
		\end{center}
	\end{figure}
	
	In this section, we study the temporal evolution of the Vorono{\"\i} volumes. Figure \ref{fig:fig8}(a) shows the temporal autocorrelation function $C_{V}(\tau)$ of the associated Vorono{\"\i} volumes for three types of swimmers at a fixed $\Psi=1$, which is defined in (\ref{AutocorreDef}) as the correlation coefficient between the Vorono{\"\i} volumes at two different time as a function of the time lag $\tau$:
	\begin{equation}
		C_{V}(\tau)=\frac{\left \langle V'(t)V'(t+\tau) \right \rangle_{o}}{\left \{\left \langle V'(t)^2 \right \rangle_{o}\left \langle V'(t+\tau)^2 \right \rangle_{o}\right \}^{1/2}},
		\label{AutocorreDef}
	\end{equation}
	where $\langle \ldots \rangle_{o}$ represents the average taken over for the overall particles at a specified time step. Note that the mean value of the Vorono{\"\i} volumes at the corresponding time have been subtracted, so the fluctuating component of the Vorono{\"\i} volumes $V'=V-\left \langle V \right \rangle_{o}$ is used for the calculation of the temporal autocorrelation function.
	The results for non-active particles are also shown for comparison. 
	Compared with the non-active particles, the $C_{V}(\tau)$ of the slow and moderate swimmers decreases relatively slowly, reflecting the trapping of particles by turbulence prolonged the time needed for the decorrelation of the Vorono{\"\i} volumes.
	We observe that, for fast swimmers, $C_V$ decorrelates much slower as compared to the slow swimmers, which is related to the result that fast swimmers have a high possibility to accumulate as shown in Figure~\ref{fig:fig4}. 
	
	Figure \ref{fig:fig8}(b) shows the decorrelation time scales of the Vorono{\"\i} volume ($\tau_V$) as a function of the stability number ($\Psi$) for fast, moderate, slow swimmers and non-active particles.
	The decorrelation time for Vorono{\"\i} volume $\tau_V$ is defined to be the time when the autocorrelation function decreased to 0.5, i.e. $C_V(\tau_V)=0.5$.
	We remark that the spatial distributions of non-active particles are independent of $\Psi$ since they simply follow the fluid velocity at the particle position. 
	We observe that $\tau_V$ for all three types of swimmers shows non-monotonic dependence on $\Psi$, with a peak around $\Psi=1\sim2.5$. We remark the decorrelation process is slow at $\Psi=16$, with $\tau_V$ approximating the peak value. The stable upward swimming direction that helps maintain the spatial distribution may partially explain this large $\tau_V$ for particles at large $\Psi$.  Using the results of non-active particles as a baseline, we can study the swimming velocity effects on the decorrelation time. The decorrelation time ($\tau_V$) for fast swimmers is the largest among the three types of swimmers at large $\Psi$, while is the smallest at small $\Psi$. In the gravity torque dominated region ($\Psi\gg1$), the uniformly stable upward swimming direction combined with a larger swimming velocity $\Phi$ (relatively smaller fluid velocity effects) results in a more stable spatial distribution and increasing decorrelation time. In contrast, in the turbulence torque dominated region ($\Psi\ll1$), the random swimming direction combined with a larger swimming velocity $\Phi$ facilitates the change of distance between particles and accelerates the decorrelation process.      
	
	\begin{figure}
		\begin{center}
			\includegraphics[width=1\columnwidth,clip]{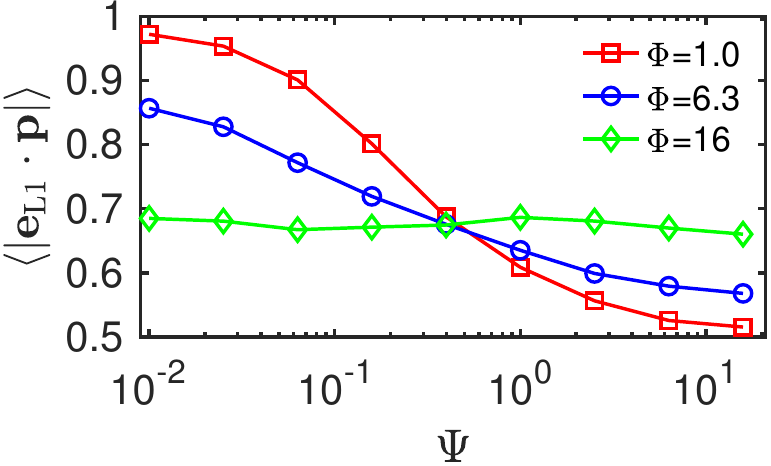}
			\caption{Alignments of particles with the Lagrangian stretching direction $\boldsymbol{e}_{L1}$ as a function of the stability number $\Psi$ for slow, moderate, and fast swimmers at $\alpha=20$. Particles with greater swimming velocity exhibit stronger alignment when $\Psi\gg1$, whereas the opposite trend is observed when $\Psi\ll1$. }
			\label{fig:fig9}        
		\end{center}
	\end{figure}

	\subsection{Alignment with Lagrangian fluid stretching}\label{sec:lag}
	
	The quantitative measurement of this Lagrangian alignment is obtained via the ensemble and temporal average of the cosine of the angle between the particle swimming direction and the Lagrangian stretching direction, $\left \langle|\boldsymbol{e}_{L1}\cdot \boldsymbol{p}|\right \rangle$.
	$\left \langle|\boldsymbol{e}_{L1}\cdot \boldsymbol{p}|\right \rangle$ takes the value 1 in the case of a perfect parallel alignment along the $\boldsymbol{e}_{L1}$ direction, 0 in the perfect perpendicular case, and 0.5 if the particles point randomly with respect to $\boldsymbol{e}_{L1}$.   
	
	We look at the mean Lagrangian alignment of particles of different stability numbers $\Psi$ for three different swimming numbers $\Phi$. Figure~\ref{fig:fig9} shows the measurements for slow ($\Phi=1$), moderate ($\Phi=6.3$) and fast swimmers ($\Phi=16$). Slow swimmers with weak gyrotaxis ($\Psi\ll1$) approach perfect alignment with the Lagrangian stretching direction, with $\left \langle|\boldsymbol{e}_{L1}\cdot \boldsymbol{p}|\right \rangle$ being close to 1. This strong alignment is consistent with the results of passive rods \cite{ni2014alignment}, reflecting the alignment behavior of slow swimmers shows only small deviations from that of non-active particles.   
	In contrast, when $\Psi\gg1$, $\left \langle|\boldsymbol{e}_{L1}\cdot \boldsymbol{p}|\right \rangle$ of slow swimmers reaches 0.5, implying that the particle swimming direction and the Lagrangian stretching direction point randomly relative to each other. This is consistent with our expectations since the influence of fluid stretching is almost negligible compared with the dominated gravity when $\Psi\gg1$. We now consider the effect of the swimming number $\Phi$ on the Lagrangian alignment, using the alignment of slow swimmers as a baseline. The dependence of alignment on $\Phi$ shows opposite trends in the turbulence ($\Psi\ll1$) and the gravity torque ($\Psi\gg1$) dominated regions. 
	The alignment decreases with the swimming velocity in the small $\Psi$ regime, whereas it becomes stronger in the large $\Psi$ regime as $\Phi$ increases.

\begin{figure}
		\begin{center}
			\includegraphics[width=1\columnwidth]{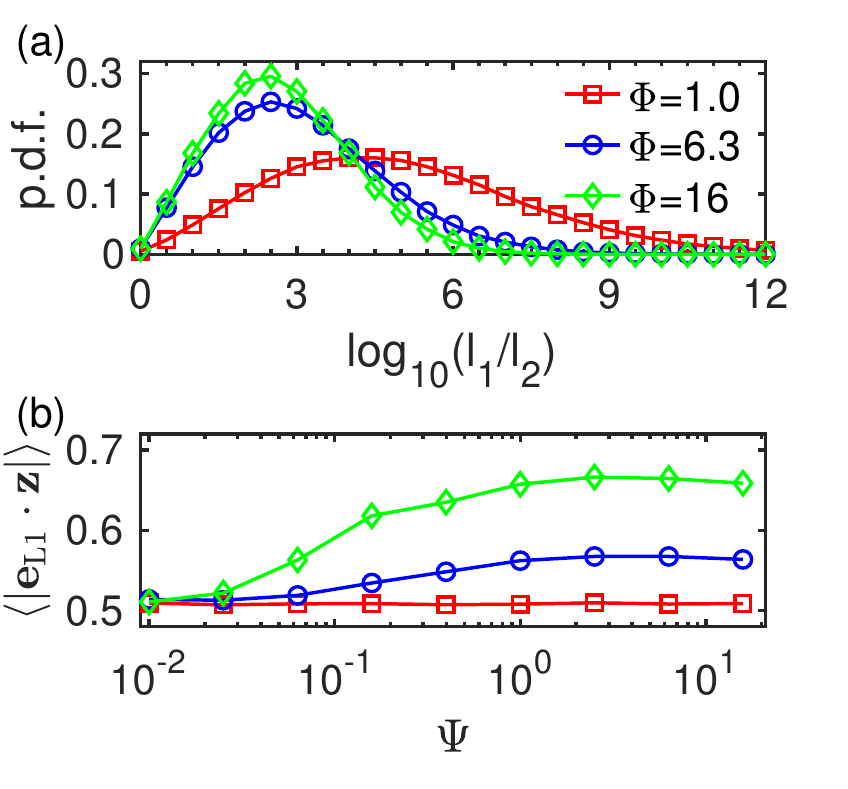} 
			\caption{(a) The p.d.f.s of the geometrical ratios $l_1/l_2$ for slow, moderate, and fast swimmers at $\Psi=0.01$. The geometrical ratio p.d.f.s are qualitatively similar for $\Psi$ in the range of [0.01,16], thus only the results at $\Psi=0.01$ are plotted here. For slow swimmers, the geometrical ratio tends to be more extreme, reflecting a stronger Lagrangian stretching effect. (b) The mean of the absolute value of the vertical component of the Lagrangian stretching direction. }
			\label{fig:fig10}       
		\end{center}
	\end{figure}

	In order to make sense of the dependence on $\Phi$ when $\Psi\ll1$, we measure the principal axes of length $l_i=\sqrt{\Lambda_i} (i=1, 2, 3)$ of an ellipsoid deformed from a sphere due to Lagrangian stretching. 
	We quantitatively measure the Lagrangian stretching strength via the p.d.f.s of the geometrical ratio  $l_1/l_2$ for three types of swimmers, see Figure \ref{fig:fig10}(a). An important feature of the p.d.f. of the slow swimmer is that the highest probability occurs at the largest $l_1/l_2$ among three swimmers. With increasing swimming velocity, the geometrical ratio has a higher probability to be in the small range. 
	It has been found that the alignment between $\boldsymbol{p}$ and $\boldsymbol{e}_{L1}$ increases monotonically with the geometrical aspect ratio $l_1/l_2$ from a statistical perspective\cite{ni2014alignment}. Therefore, the geometrical aspect ratio $l_1/l_2$, to some extent, can reflect the strength of this Lagrangian stretching. To understand the weakened Lagrangian stretching due to swimming, we conjecture the following physical image.
	The Lagrangian velocity gradient experienced by particles with a large swimming velocity decorrelates faster than that of non-active particles. As a result, the typical time scale for the process of Lagrangian stretching becomes shorter. Therefore, the alignment with the Lagrangian stretching direction decreases as the swimming velocity increases at $\Psi\ll1$, as shown in Figure~\ref{fig:fig9}.
	
	The analysis above is for $\Psi\ll1$. As we have examined, three swimmers give qualitatively similar geometrical ratio ($l_1/l_2$) p.d.f.s for $\Psi\in[0.01, 16]$. These p.d.f.s for other $\Psi$ are not shown in Figure~\ref{fig:fig10}(a), since these curves nearly collapse with the shown one. Therefore, the conclusion that fast swimmers experience weaker Lagrangian stretching is robust in the large $\Psi$ range. However, we see in Figure \ref{fig:fig9} that fast swimmers under strong gyrotaxis ($\Psi\gg1$) show better alignment than moderate and slow swimmers, although they have encountered weaker fluid stretching. This deviates from the usual expectations. Figure~\ref{fig:fig10}(b) shows the mean of the vertical component of $\boldsymbol{e}_{L1}$, $\left \langle|\boldsymbol{e}_{L1}\cdot \boldsymbol{z}|\right \rangle$, as a function of $\Psi$ for different $\Phi$.
	It turns out that the Lagrangian stretching direction can be biased by gravity. The Lagrangian stretching direction of fast swimmers shows the largest extent to align with the vertical direction. Under the strong gyrotaxis effect, the swimming direction of particles also approaches the vertically upward direction $\boldsymbol{z}$. Thus, the counterintuitive strong alignment for fast swimmers at large $\Psi$ can be understood.  
	However, the bias of $\boldsymbol{e}_{L1}$ in the vertical direction is not revealed.
	The Lagrangian stretching direction is obtained by integrating the velocity gradient $\boldsymbol{A}(\boldsymbol{x},t)$ (\ref{deformation}), which does not include the gravity effect explicitly. One possible reason could be the biased sampling of particle positions $\boldsymbol{x}$ in the flow, which was found to be upwelling or downwelling regions \citep{gustavsson2016preferential,lovecchio2019chain}. Even if there is no preferential accumulation when $\Psi\gg1$, the uniformly upward swimming direction of particles may prolong their time staying in structures with large vertical velocity. 
	In these regions the flow may tend to stretch particles in the vertical direction, thus causing the bias of $\boldsymbol{e}_{L1}$ in the vertical direction. More studies are needed to fully understand this issue.

	\section{Conclusions\label{sec:conclusions}}
	In this paper, we consider simulations of the dynamics of elongated gyrotactic swimmers in turbulent flows. We carry out the study of clustering of gyrotactic swimmers, using the three-dimensional Vorono{\"\i} analysis. Three different types of swimmers (fast, moderate, and slow) of different $\Psi$ are covered. At small and intermediate $\Psi$, a higher probability of finding depleted regions (large Vorono{\"\i} volumes) and concentrated regions (small Vorono{\"\i} volumes) is identified for active particles, implying the clustering behavior. We use the standard deviations of the normalized Vorono{\"\i} volumes to measure the clustering extent quantitatively. At intermediate $\Psi$, fast swimmers show stronger clustering than the other two types of swimmers, which is consistent with the trend based on the radial distribution function analysis. While the gyrotaxis-dominated ($\Psi\gg1$) swimmers exhibit no clustering, non-gyrotactic particles ($\Psi\ll1$) are amenable to concentrate preferentially, with a non-monotonic dependence on $\Phi$.  
	
	Next, we study the connections between the Vorono{\"\i} volume and the local flow properties from the spatial and time perspective. By comparing the p.d.f. of the overall particles and that of particles belonging to clusters and voids, the locations that overall, aggregated, and sparse particles preferentially sample are further distinguished. Clusters and voids preferentially form at locations with more extreme vertical velocity compared with the overall preferential sampling. The preferential vertical flow regions of clusters and voids are roughly the opposite. From the Lagrangian autocorrelation of the Vorono{\"\i} volumes, the lifetime of the clusters is examined. The clustering of fast swimmers lasts longer than that of moderate and slow swimmers at large $\Psi$, but the opposite dependence on swimming velocity emerges at small $\Psi$. In the future, more insights may be gained by following individual clusters of gyrotactic swimmers, focusing on how they are formed and destroyed by turbulent fluctuations.
	
	Finally, we study their Lagrangian alignment, focusing on the Lagrangian stretching directions. We have demonstrated how the motility can weaken this alignment, by visualizing stretching as the process of deforming a sphere into an ellipsoid. With faster velocity, the deformation is reduced, reflecting a weakened Lagrangian stretching. However, when gravity effects become more dominated, the Lagrangian alignment shows a counterintuitive increasing trend with a particle moving velocity. We found, under this case, the Lagrangian stretching direction can also be biased by gravity. Gravity biases both the particle swimming direction and the Lagrangian stretching direction in the vertical direction. Therefore, the remarkable alignment trend with the particle moving velocity at large $\Psi$ can be understood.
	
	\textit{Acknowledgments} 
	We thank Enrico Calzavarini and Z. Wang for their comments on the manuscript.
	This work is financially supported by the Natural Science Foundation of China under Grant No.~11988102, 91852202, the National Training Program of Innovation and Entrepreneurship of China for Undergraduates (202110003017), and Tencent Foundation through the XPLORER PRIZE.
	
	%
	
\end{document}